\documentclass[preprint, dvips]{imsart}

\RequirePackage[OT1]{fontenc}
\RequirePackage{amsthm,amsmath}
\RequirePackage[numbers]{natbib}
\RequirePackage[colorlinks,citecolor=blue,urlcolor=blue]{hyperref}
\RequirePackage{graphicx, graphics, epsfig, amsfonts, amssymb}
\RequirePackage[parfill]{parskip}
\RequirePackage[small,compact]{titlesec}

\startlocaldefs
\numberwithin{equation}{section}
\theoremstyle{plain}
\endlocaldefs

\begin{document}

\begin{frontmatter}
\title{Terahertz Generation and Amplification in Graphene Nanoribbons in Multi-frequency Electric Fields}
\runtitle{Terahertz generation and amplification in graphene nanoribbons}

\begin{aug}
\author{\fnms{M.} \snm{Rabiu}\thanksref{}\ead[label=e1]{rabpeace10gh@gmail.com}}
\address{Department of Applied Physics, Faculty of Applied Sciences, University\\
for Development Studies, Navrongo Campus, Ghana.\\
\printead{e1}}

\author{\fnms{S. Y.} \snm{Mensah}\thanksref{}\ead[label=e2]{profsymensah@yahoo.co.uk}}
\and
\author{\fnms{S. S.} \snm{Abukari}\thanksref{}\ead[label=e3]{asseidu75@yahoo.com}}
\address{Department of Physics, Laser and Fiber Optics Center, University \\ 
of Cape Coast, Cape Coast, Ghana.\\
\printead{e2,e3}}

\runauthor{Rabiu Musah et al.}
\end{aug}

\begin{abstract}
We study theoretically a multi-frequency response of electrons in confined graphene subject to dc-ac driven fields. We explore the possibility for using graphene nanoribbons (GNRs) to generate and amplify terahertz (THz) radiations in electric field domainless regime. We discover two main important schemes of generation; when the frequencies are commensurate, THz generation is due to wave mixing and when they are non-commensurate, a single strong field suppresses space charge instability and any weak signals can get amplified. The use of graphene as a best substitute for semiconductor nanoelectronic devices is suggested.
\end{abstract}

\begin{keyword}
\kwd{Graphene}
\kwd{Current density}
\kwd{Bessel function}
\kwd{Terahertz}
\end{keyword}

\received{\smonth{2} \syear{2012}}
\end{frontmatter}

\section{Introduction}
Graphene is a monolayer of one atom thick with fascinating carrier transport properties. Especially, its high current density and high carrier mobility of 44000$cm^2V^{-1}s^{-1}$ \cite{RSShishir}. But attempts to utilize these unique properties in graphene devices is posing some difficulties. The limitation is probably due to several factors including; lack of bandgap in graphene sheets, edge defects, disorder, among others. To overcome some of these obstacles, the dimention of graphene sheets has to be reduced or the geometry altered. After all, new physics (quantization) emerge when dimensions of materials are reduced. An infinite 2D graphene  could become 1D + quantization along one other direction opening a gap. The resulting material is known as graphene nanoribbon (GNR). Depending on the nature of the edges, one can get two symmetry groups from this GNR; armchair graphene nanoribbon (aGNR) or zigzag graphene nanoribbon (zGNR). Electron dynamics of both aGNR and zGNR have different electronic properties, mostly due to the berry phase and pseudospin \cite{KSasaki}. Edge states have significant contribution to graphene properties, because in a nanometer size ribbon, massless Dirac fermions can reach the edges within a femto-second before encountering any other lattice effects, like electron-electron interaction, electron-phonon interaction, etc.

In this paper, we study the phenomenon of generating frequencies in the terahertz (THz) range. The development of sources and sensors emitting and detecting electromagnetic waves in the terahertz regime has been the subject of interest for some time now. And  holds great promise for graphene based THz metamaterials, optoelectronic devices, THz lasers, fast switching mechanisms, spectroscopy, wireless communication \cite{DDragoman0}. Recently, THz generations are studied in graphene by resonance tunneling-like configuration \cite{DDrogmanM}, by tunable plasmon excitations and light-plasmon coupling \cite{LongJu} and by optical pumping of graphene \cite{VRyzhii, AADubinov, TOtsuji}. Bloch oscillations up to 10 THz can be generated in periodic graphene structures \cite{DDrogman}. Today, semiconductor superlattices are used as sources for THz radiation and detection. However, GNRs are better candidates because of their low dimensionality, striking electronic properties and the possibility of controlling these properties via applied gate voltage. Graphene is also relatively easy to fabricate in laboratory.

The physical mechanism governing THz generation in graphene, when subject to applied electric field, can be understood in terms of ballistic trajectories of electrons in the quasi-momentum space. When graphene is subject to an electric field, ballistic acceleration of charge carriers generates to-and-fro motion of the whole distribution function, which varies from zero to several electron volts. It is a collective motion of these charges that manifest THz oscillation of carriers in graphene. The highly nonlinearity of graphene as a carbon allotrope and the fact that it has non-parabolic energy spectrum can also account for THz production in the material. This last effect is more applicable if the frequencies are commensurate. When two or more commensurate frequencies interfere in a region they could result in creation of fields with zero frequencies, i.e static fields. These bias fields are responsible for Bloch oscillations at THz frequencies \cite{THyart}.

The remaining of this paper is organized as follows; In section \ref{sec:CurrentDensity}, we introduce the current density of aGNR and zGNR and imposed certain conditions to reduce the equations to simple forms appropriate for our systems under discussion. By limiting the harmonics fields to only two terms, we deduce I-V characteristic equations for describing THz generations in section  \ref{sec:CurrentGeneration}. The equations obtained in the preceding section are plotted and discussed in section \ref{sec:Resultsdiscussions}, with conclusion and some recommendations for future applications in section \ref{Conclusions}.

\section{Current density equation} \label{sec:CurrentDensity}
For detailed calculation of current density for GNR, see our recent paper \cite{NotYet}. To avoid book keeping, we state without repetition of the proof, the relation for the sheet current density when graphene is subject to an external field of the form $E(t) = E_0 + \sum_j E_jcos(\omega_jt + \alpha_j)$ as
\begin{equation}
	j(t) = i\sum_{r = 1}^{\infty}j_{0r}\left[ \sum_{n_j,\,\nu_j=-\infty}^{\infty}\prod_{j=1}^nJ_{n_j}(r\beta_j)J_{n_j-\nu_j}(r\beta_j)\frac{e^{i\nu_j\omega_jt + i\nu_j\alpha_j}}{1+i\tau(r\beta_0 + n_j\omega_j)} + c.c\right].\label{eq:ja0}
\end{equation}
Where $j_{0r}$ is the peak current density, $J_n(\beta)$ is a Bessel function of order $n$ and argument $\beta = el\tau E/\hbar\omega$ and $\beta_0 = elE_0/\hbar$. For the rest of this paper we will consider a maximum of two harmonic frequencies $\omega_1$ and $\omega_2$. Because $n = 2$, the above equation will look like 
\begin{eqnarray}
	j(t) &=& i\sum_{r = 1}^{\infty}j_{0r} \sum_{n_1\,n_2,\,\nu_1\,\nu_2=-\infty}^{\infty}\frac{J_{n_1}(r\beta_1)J_{n_1-\nu_1}(r\beta_1)J_{n_1}(r\beta_2)J_{n_2-\nu_2}(r\beta_2)}{1+i\tau(r\beta_0 + n_1\omega_1 + n_2\omega_2)}\nonumber\\
			 &&\times e^{i(\nu_1\omega_1 + \nu_2\omega_2)t + i\nu_1\alpha_1  + i\nu_2\alpha_2} + c.c,\label{eq:ja}
\end{eqnarray}
where 
\[
	j_{0r} = \frac{2g_sg_ve\gamma_0}{\pi l\hbar}\Delta\theta\sum_{s=1}^{n}r\mathcal{E}_{rs}f_{rs}
\]
and $g_s$, $g_v$ are the spin and valley degeneracies. $l=\sqrt{3}a/2$, $\Delta\theta = \pi s/(n+1)$ for aGNR and $l=a/2$ and $\Delta\theta = \pi (s+1/2)/(n+1)$ for zGNR.

\section{Current generation and THz amplification}\label{sec:CurrentGeneration} 
When the electric field is applied to graphene, there naturally arises two schemes of generation and amplification. (a) unbiased, $E_0 = 0$ at even harmonics with commensurate frequencies and (b) biased, $E \neq 0$ at non-commensurate frequencies. In the following sections, we study both scenarios in details.

\subsection{Commensurate frequencies}
If one averages out Eq.\eqref{eq:ja} over a period of the GNR in both sides of the equation, we get $\langle j(t) \rangle = j$ and in the right hand side a delta function emerges which ensures that $\nu_1 = -\frac{\omega_2}{\omega_1}\nu_1$. Further, we consider an applied field consisting of purely periodic multiple harmonic frequencies, i.e $\omega_2=\mu\omega_1$, $\omega_1=\Omega$ with $\mu$ being an integer or fraction. Taking the sum over $n_2$ after linearizing with respect to $E_2$ in the weak field $\beta_2 << 1$ limit. This restricts the order $n_2$ of the Bessel function to take only small values $\pm 1$. The real part of the current density then becomes
\begin{equation}
j=i\sum_{r=1}^{\infty}j_{0,r}\sum_{n_1=-\infty}^{\infty}\frac{J_0(r\beta_2)\sum_{n_2=\pm1}J_{n_2}(r\beta_2)J_{n_1}(r\beta_1)J_{n_1 + \mu n_2}(r\beta_1)}{1+i\tau(r\beta_0 + [n_1 + \mu n_2]\Omega)}e^{in_2\alpha}. \label{eq:ja1}
\end{equation}

The integer $\mu$ can take even or odd values. Odd integer values will yield an imaginary current density. This means that odd harmonics are not very interesting for THz generations but can still exhibit NDC \cite{YuARamanov}. Substituting $J_0(r\beta_2)J_{+1}(r\beta_2) \sim r\beta_2/2$ in the preceding equation we arrived at 
\begin{eqnarray}
	j &=& j_{0} \frac{el\tau}{\mu\hbar\Omega}E_2cos\alpha\sum_{r=1}^{\infty}\sum_{s=1}^{\mathcal{N}}r^2\mathcal{E}_{rs}f_{rs}\nonumber\\
	  &&\times \sum_{n=-\infty}^{\infty}\left[\frac{J_{n}(r\beta_1)J_{n + \mu}(r\beta_1)}{1 + i\tau(r\beta_0 + [n + \mu]\Omega)} - \frac{J_{n}(r\beta_1)J_{n - \mu}(r\beta_1)}{1 + i\tau(r\beta_0 + [n - \mu]\Omega)}\right].\label{eq:ja2}
\end{eqnarray}
We have obtained this equation due to wave mixing of commensurate frequencies. Direct Bloch oscillations could still be induced even for zero static field, but for $\mu$ equal to half-integers. We do not consider that here, so we fix $\beta_0 = 0$. In fact, it is not hard to bring the real part of Eq.\eqref{eq:ja2} to the form in \cite{SSAbukari3}, i.e 
\begin{equation}
	j = j_{0} \frac{el\tau^2}{\mu\hbar}E_2cos\alpha\sum_{r=1}^{\infty}\sum_{s=1}^{\mathcal{N}}r^2\mathcal{E}_{rs}f_{rs}\left[\sum_{n=-\infty}^{\infty}\frac{nJ_{n}(r\beta_1)J_{n - \mu}(r\beta_1)}{1 + (n\Omega\tau)^2}\right].\label{eq:ja3}
\end{equation}
However, the advantage of our equation over \cite{SSAbukari3} is that $\mu$ does not have to be $2$ only.

\subsection{Non-commensurate frequencies}
If the frequencies are not commensurate, then the condition $\nu_1 + \mu\nu_2 = 0$ is lifted and $\omega_2 \neq \nu\omega_1$. However, it is still possible to amplify frequencies in the THz domain if one considers $\omega_1$ and $\omega_2$ as pump and probe frequencies respectively. Both frequencies belong to THz range, i.e $\omega_1\tau  \gtrsim 1$ and $\omega_2\tau  \gtrsim 1$. This approach has been adopted in \cite{THyart0, THyart} for superlattices and in experiment for generating THz using resonance tunneling-like configuration in graphene \cite{DDrogmanM} and amplifying small frequencies in epitaxially grown graphene heterostructures \cite{TOtsuji}. Here, the THz is generated because the pump field excites electrons, the space charge instability is suppressed by the strong (pump) field while the week probe signals get amplified.

The real part of Eq.\eqref{eq:ja} for the two frequencies which are not related ($\alpha_{1,2}=0$) takes the form
\begin{equation}
	j = i\sum_{r = 1}^{\infty}j_{0r}\left[ \sum_{n_1\,n_2=-\infty}^{\infty}J^2_{n_1}(r\beta_1)J^2_{n_2}(r\beta_2)\frac{r\beta_0\tau + n_1\omega_1\tau + n_2\omega_2\tau}{1+[\tau(r\beta_0 + n_1\omega_1 + n_2\omega_2)]^2}\right],\label{eq:ja4}
\end{equation}
where we have assume that $\nu_1=\nu_2=0$.

Most often, we shall be using the ratio $E_{1,2}/E_{cr}$ which is defined as 
\begin{equation}
	 \frac{E_{1,2}}{E_{cr}} = \beta_{1,2}\omega_{1,2}\tau, \quad \mbox{ with } \quad E_{cr} = \frac{\hbar}{el\tau} 
\end{equation} 
instead of just $\beta_{1,2}$.

\section{Results and Discussion} \label{sec:Resultsdiscussions}
\begin{figure}[thb!]
	\centering{\includegraphics[scale=0.75]{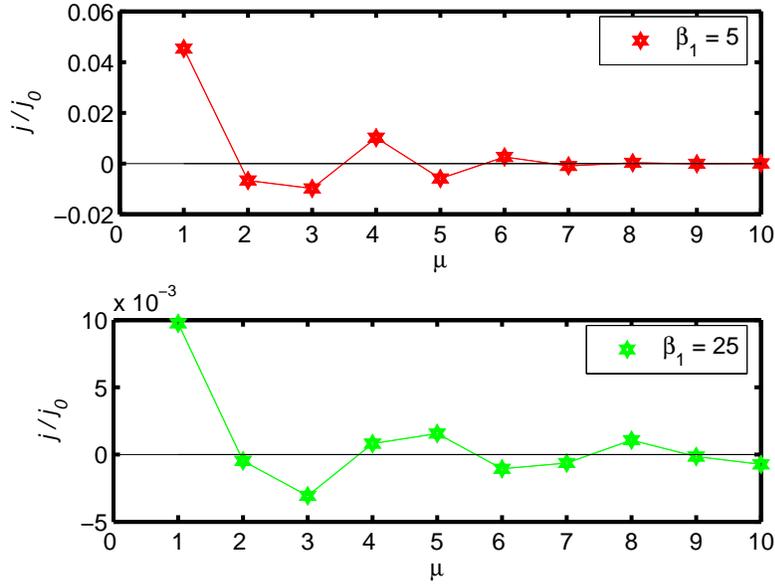}}
	\caption{Non-linear current density with harmonic index, $\mu$. $cos(\alpha) = 1$}
	\label{fig:thz0}
\end{figure}

\begin{figure}[thb!]
	\centering{\includegraphics[width=4.5in,height=2.35in]{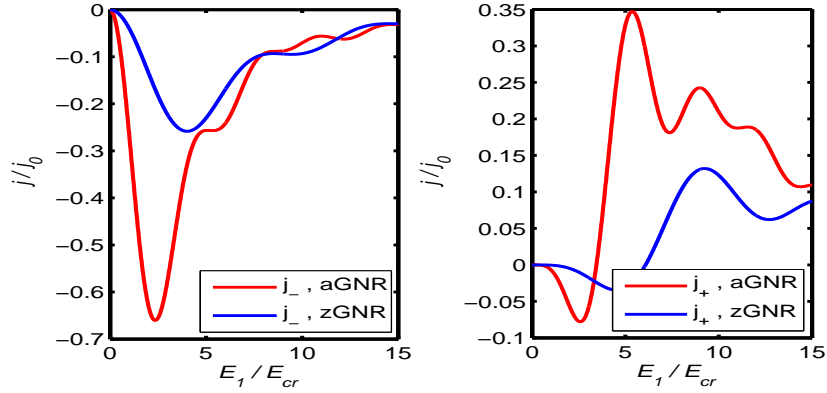}}
	\caption{THz current density against ac field amplitudes for armchair and zigzag graphene nanoribbons. (Left) second harmonic, $\mu = 2$ and (right) fourth harmonic $\mu = 4$. $cos(\alpha) = 1$.}
	\label{fig:thzz1}
\end{figure}

In Fig.\ref{fig:thz0}, we have demonstrated dependence of non-linear current density on the harmonic index $\mu$ for week (top) and strong (buttom) ac amplitudes. In both cases, some $\mu$ values give positive ($j_-$) and negative ($j_+$) current density. The series resulting in $j_+$ yields Bloch oscillations of the current density that decays faster to $j=0$ from above as in Fig.\ref{fig:thzz1} (right). The series that results in $j_-$ produces Bloch oscillations that is asymptote to $j = 0$ from below as in Fig.\ref{fig:thzz1} (left). This means electronic oscillations persist for some time before dying off and thus the $\mu$-series that give $j_-$ is the better option for THz production. They are two sub-categories of $j_{\pm}$. $j_+$: $\mu_{odd} = 1, 9, \ldots$, $\mu_{even} = 4, 6, 8, 10\ldots$ at low ac amplitudes and $\mu_{odd} = 1, 5, \ldots$, $\mu_{even} = 4, 8, \ldots$ at high amplitudes. $j_-$: $\mu_{odd} = 3, 5, 7, \ldots$, $\mu_{even} = 2,\ldots$ at low ac amplitude and $\mu_{odd} = 3, 7, 9, \ldots$, $\mu_{even} = 2, 6, 10, \ldots$ at high amplitudes. From the graphs, we found that $\mu_{even}= 2, 4$ and $\mu_{odd} = 1, 3$ are robust against the ac field amplitudes, $E_1$, $E_2$.

The graphs in Fig.\ref{fig:thzz1} show the behavior of terahertz current density on the ac field amplitudes for armchair and zigzag graphene nanoribbons at $\mu = 2, 4$. For low $E_1/E_{cr}$ values at $\Omega\tau = 2$, the absolute current steadily increases to maximum before falling. The curve then begins to oscillate above certain ac threshold amplitude, $E_{1,min}$. Under our default parameters, $E_{1, min} = 2E_{cr}$ for aGNR and $E_{1, min} = 2\sqrt{3}E_{cr}$ for zGNR. This oscillations are predicted to lie within THz range \cite{SSAbukari3}.

\begin{figure}[thb!]
	\centering{\includegraphics[width=4.0in,height=2.7in]{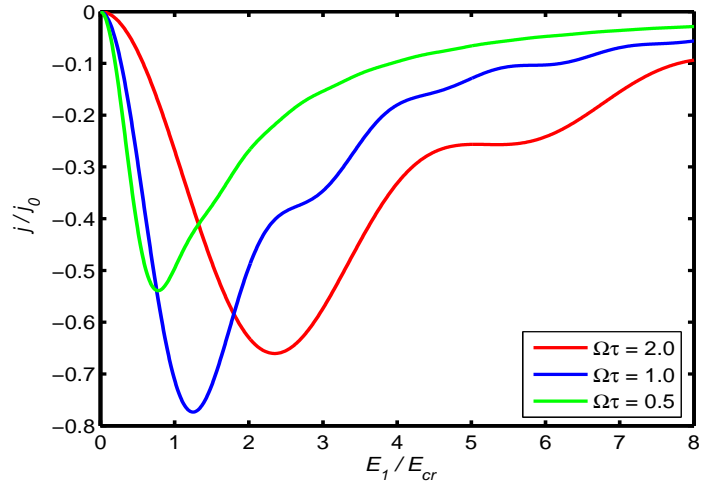}}
	\caption{THz current density against ac field amplitudes for armchair and zigzag nanoribbon at various $\Omega\tau$ values. $cos(\alpha) = 1$}
	\label{fig:thzmu2}
\end{figure}

We also studied in Fig.\ref{fig:thzmu2} the combined effect of Thz current with reduced ac amplitudes and frequencies. Oscillations disappear when $\Omega\tau << 1$. This region is not feasible for THZ generation. However, for $\Omega\tau \geq 1$ oscillaions are more pronounced and a graphene device under this condition can operate effectively to produce THz frequencies. 

\begin{figure}[thb!]
	\centering{\includegraphics[width=3.7in,height=2.75in]{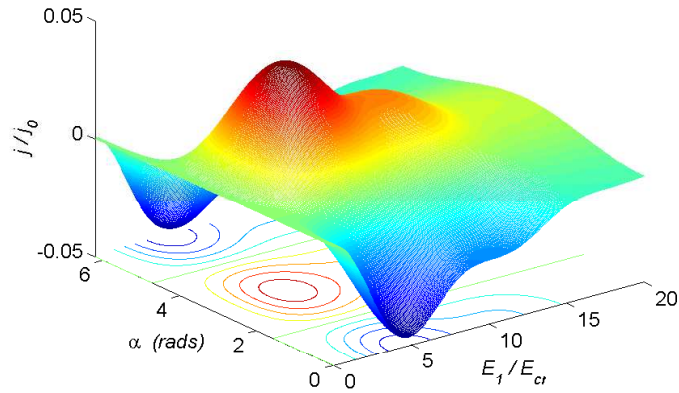}}
	\caption{Three dimensional plot showing THz current oscillations at varying ac field and phase difference.}
	\label{fig:ThreeD_alpha}
\end{figure}

Within the scheme of commensurability of frequencies, we have finally demonstrate the behavior of THz current on both phase difference and ac amplitudes in Fig.\ref{fig:ThreeD_alpha}. There are some phase differences where there is no oscillations, this is indicated on the contour as straight lines. Interestingly, current peaks when the two ac fields are out of phase, i.e $\alpha = \pi$ and $E_1 = 5E_{cr}$.

\begin{figure}[thb!]
	\centering{\includegraphics[width=4.7in,height=2.35in]{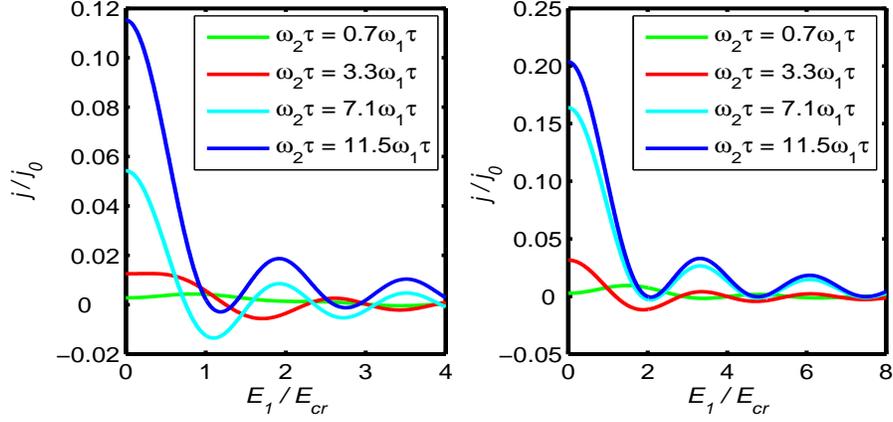}}
	\caption{THz current oscillations at non-commensurate frequencies. $E_0 = 2.5E_{cr}$.}
	\label{fig:inthz1}
\end{figure}

Now, we turn to the case of non-commensurate frequencies. In Fig.\ref{fig:inthz1}, we plot THz current with ac amplitudes in the presence of another stronger ac amplitude but with weak frequency and biased static field. It is this weak frequency that excites electrons after the electric charge instability is suppressed by the strong ac field amplitude. The excited electrons Bloch oscillates at rather magnified frequency within the THz range. However, the nature of oscillations in the figure is different from what we have seen earlier. Here, the curve falls quickly and then begins to oscillate almost immediately.

\begin{figure}[thb!]
	\centering{\includegraphics[scale=0.55]{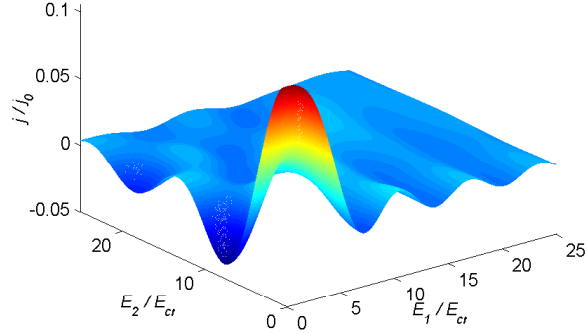}}
	\caption{Three dimensional plot showing THz current oscillations at varying ac fields. $\omega_2 = 0.7\omega_1$, $E_0 = 2.5E_{cr}$}
	\label{fig:inthzmu3D1}
\end{figure}

The graph in Fig.\ref{fig:inthzmu3D1} also shows effects of both the ac fields $E_1$, $E_2$ on THz current.

\section{Conclusion}\label{Conclusions}
We have studied theoretically, two schemes of THz production; generation and amplification at commensurate and non-commensurate frequencies respectively. For commensurable frequencies, we discovered two harmonic series, $\mu_{odd} = 2, 4, ...$ and $\mu_{even} = 1, 3, ...$ which are robust against high ac field amplitudes and at which values production is suitable. Generation at the commensurate frequencies is due to wave mixing at zero bias field. For incommensurable frequencies, there is THz amplification of small ac signals. A strong ac amplitude is neccessary to suppress charge instability at non-zero bias field. 



\end{document}